\documentclass[iop, twocolumn, tighten]{emulateapj}
\usepackage{natbib} 
\usepackage{color}

\newcommand{\Nsampel}{N_\mathrm{samp}}
\newcommand{\NN}{ 61, 56, 39, and 13 }

\newcommand{\Nsample}{N_\mathrm{cl}}
\newcommand{\tR}{\tilde{R}}

\newcommand{\Sx}{S_\mathrm{X}}

\newcommand{\muA}{\mu_A}
\newcommand{\muB}{\mu_B}
\newcommand{\muAgas}{\mu_{g,A}}
\newcommand{\muBgas}{\mu_{g,B}}

\newcommand{\rmin}{r_{\mathrm{min}}}

\slugcomment{}
\shortauthors{Kawahara}
\shorttitle{}

\begin{document}
\title{The axis ratio distribution of X-ray clusters observed by XMM-Newton}

\author{Hajime Kawahara\altaffilmark{1,2}} 
\altaffiltext{1}{Department of Physics, Tokyo Metropolitan University,
  Hachioji, Tokyo 192-0397, Japan}
\altaffiltext{2}{Department of Physics, The University of Tokyo, 
Tokyo 113-0033, Japan}
\email{kawa\_h@tmu.ac.jp}

\begin{abstract}
We derive the axis ratio distribution of X-ray clusters using the XMM-Newton catalogue \citep{2008A&A...478..615S}. By fitting the contour lines of the X-ray image by ellipses, we confirm the X-ray distribution is well approximated by the elliptic distribution with a constant axis ratio and direction. We construct a simple model describing the axis ratio of the X-ray gas assuming the hydrostatic equilibrium embedded in the triaxial dark matter halo model proposed by \citet{2002ApJ...574..538J}  and the hydrostatic equilibrium. We find that the observed probability density function of the axis ratio is consistent with this model prediction. 
\end{abstract}

\section{Introduction}

 Galaxy clusters have been playing an important role in the determination of cosmological parameters such as the matter density, the amplitude of the initial fluctuation, the Hubble constant. Upcoming surveys of galaxy clusters via optical, X-ray, lensing and the Sunyaev-Zel'dovich effect will enable one to probe new aspects of cosmology such as the nature of the dark energy \citep[e.g.][]{2006astro.ph..9591A} and the initial non-Gaussianity. Therefore, the physical modeling of galaxy clusters is important to properly interpret the cluster data. In particular, non-sphericity adds significant uncertainty to various cosmological applications of the clusters \citep[e.g.][]{2003ApJ...599....7O,2004ApJ...609...50D}. 

 The statistical nature of the non-sphericity has been investigated using N-body simulations by many authors \citep[e.g.][]{2002ApJ...574..538J,2005ApJ...629..781K,2006MNRAS.366.1503P,2006MNRAS.367.1781A,2007MNRAS.376..215B}.  Among them, \citet[][hereafter JS02]{2002ApJ...574..538J} proposed a triaxial model of the dark matter halo based on the detailed analysis of N-body simulations. They also provided the fitting formula of the probability distribution of the axis ratio. 

 Their phenomenological model has been applied to many studies that investigate the systematic errors of the cosmological applications with clusters. 
 The observational test for the axis ratio distribution, however, is quite limited. \citet{2009ApJ...695.1446E} analyzed 4281 clusters in the Sloan Digital Sky Survey using the gravitational lensing. By stacking all the clusters, they were merely able to conclude that the axis ratio of the projected dark matter halo $f = 0.48^{+0.14}_{-0.09}$. Due to large uncertainty, however, the measurement of the distribution of the axis ratio was difficult. Future surveys will make a significant improvement of the direct measurement of dark matter haloes with new techniques \citep[e.g.][]{2009MNRAS.400.1132H}.

On the other hand, X-ray clusters also represent the non-sphericity, which makes the large statistical error to measurement of the Hubble constant \citep[e.g][]{2008ApJ...674...11K}. The theoretical predictions of the probability density function (PDF) of the axis ratio using X-ray clusters have been considered by several authors \citep[e.g.][]{2004ApJ...617..847W, 2003ApJ...585..151L}. These are based on the assumption that the non-sphericity of X-ray clusters originated from the non-spherical gravitational potential due to the dark matter halo. It is crucial to see if the non-sphericity of X-ray clusters can be explained by the non-sphericity of the underlying dark matter halo. 
 
There are a lot of papers which measured the axis ratio of the X-ray clusters observed by Einstein  \citep[e.g.][]{1989ApJS...70..723M, 1992ApJ...400..385B, 1995ApJ...447....8M}, ROSAT \citep[e.g.][]{1996ApJ...457..565B,1997MNRAS.292..920W,2001MNRAS.320...49K} and Chandra \citep[e.g.][]{2007A&A...467..485H,2009ApJ...691.1648F}. \cite{2007MNRAS.377..883F} have constructed the theoretical prediction of the gas distribution and justified their model by comparing with 8 simulated clusters. Applying this model to 46 simulated dark matter haloes, they predicted the average and variance of the axis ratio of X-ray clusters. Comparing with ROSAT samples they have found good agreement between the predicted and observed average and variance. 


Because the XMM-Newton has the largest effective area among present X-ray satellites and its field of view is also large, one can investigate the details of the X-ray shape or isocontour lines of clusters including the radial dependence of the axis ratio and the direction of semi major axis for each cluster.  While the axis ratio constraints are not expected to be much better than have been obtained previously, it is useful to have an independent computation to serve as a check on systematic errors between different data sets and different computational methods. In this paper, we perform detailed comparisons of the PDF derived by the X-ray analysis using high quality data set of the XMM-Newton cluster catalogue \citep{2008A&A...478..615S} with the simple theoretical prediction of the PDF based on the phenomenological model of shape of underlying dark matter haloes provided by JS02.

The rest of the paper is organized as follows. In \S 2, we describe the cluster data we use and the data processing such as fitting clusters and estimating the statistical errors. In \S 3, in order to model the non-sphericity, we investigate the radial dependence of the axis ratio and the axis direction. After describing a simple model of the gas non-sphericity, we compare the non-spherical properties of X-ray clusters with the theoretical prediction. Finally, we summarize our results in \S 4.

\section{Methods}

We analyze 70 clusters in the XMM-Newton cluster catalogue \citep{2008A&A...478..615S}. The surface brightness map of each cluster is normalized by the exposure map and has the energy range of 0.5-10.0 keV band. Although XMM-Newton has three CCD imagers, 2 MOSs and pn, we simply combine the images of MOS1 and MOS2 and do not use the data of pn. Based on the light curve with the energy range of 10-15 keV, we exclude the period of " {\it flares} " due to the soft proton background from our analysis. The point sources are also excluded by the wavelet detector in the CIAO package. The images are binned in a pixel size of 1 \% of the virial radius $r_{200}$. The virial radius and mass are estimated by the relations found in the simulation, $r_{200} = 1.25 \times (1+z)^{-3/2} (T/10.0 \, \mathrm{keV})^{1/2} h^{-1} \, \mathrm{Mpc}$ and  $M_{200} = 1.5 \times (T/ 10.0 \, \mathrm{keV})^{3/2} h^{-1} 10^{15} M_{\odot} $\, \citep{1996ApJ...469..494E, 1999MNRAS.305..631A}. We adopt the averaged cluster temperature in the $1'-4'$ annulus given by \cite{2008A&A...478..615S}.

We perform the elliptical fitting according to \cite{1987MNRAS.226..747J}.  The elliptical fitting has four free parameters, the minor to major axis ratio, $q$, the direction of the long axis $\Theta$, and the center ${\bf r_c}$. In \S \ref{sec:pdf}, we compare the shape with the prediction based on the results of JS02 and they measure their PDF around the radius of $ \sim 0.3 r_{200}$. Therefore we fix the semi-major axis $R_i$ of the ellipse to $0.1$, $0.2$, $0.3$ and $0.4 \, r_{200}$ for $i=1,2,3$ and $4$. 

The background level ${\Sx}_B$ is estimated at the outer part of the surface brightness profile, $\overline{\Sx} $ around the center. We compute the average of signal to noise ratio on the ellipse as
\begin{eqnarray}
\overline{S/N} = \frac{1}{\Nsampel}  \sum_{i=1}^{\Nsampel} \frac{N(E_i)-\overline{N}_B(E_i)}{\sqrt{N(E_i)}},
\label{eq:cond}
\end{eqnarray}
where $\overline{N}_B(E_i)$ is the average count of the background estimated by ${\Sx}_B$ and the exposure map at $E_i$. we adopt the number of sampling points, $\Nsampel=360$. We use the ellipse that satisfy $\overline{S/N} > 1$.

In order to see the radial dependence of shape we independently fit the center as free parameter for each contour line). Because there is no cluster with strong double peaks  fortunately, except for RXJ 658-55 (the Bullet cluster), the best fit centers are close to the maximum surface brightness point in almost clusters. The statistical errors of $q$ and $\Theta$ due to the Poisson noise are estimated by the Monte Carlo method. We create 100 mock clusters by redistributing the photon number according to Poisson distribution. The standard deviation $\sigma_q(R_i)$ is derived by fitting the contour lines of each mock sample with the same fitting method.

\begin{figure}[!tbh]
  \centerline{\includegraphics[width=85mm]{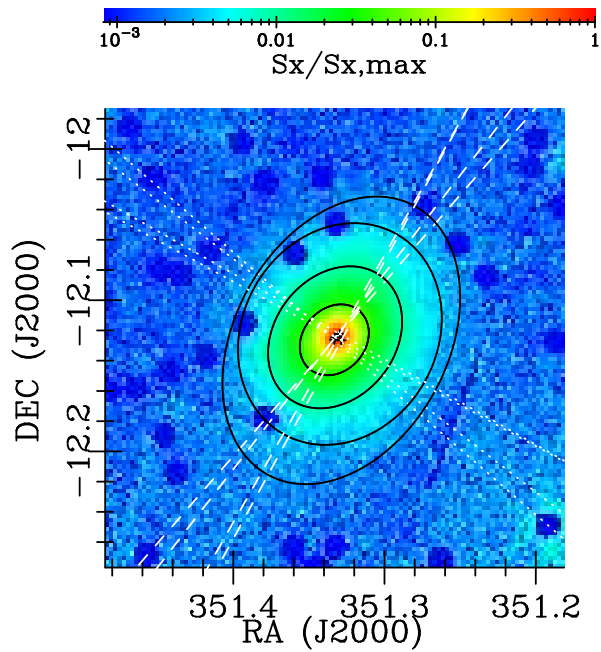}}
  \caption{An example of the elliptical fitting of the X-ray cluster (A2597). Black curves are the fitting ellipse for $R/r_{200}=0.1,0.2,0.3$ and $0.4$ by the method of \cite{1987MNRAS.226..747J}. Four white dashed and dotted lines indicates the direction of the semi major and semi minor axes, respectively. \label{fig:ex}}
\end{figure}

Figure \ref{fig:ex} displays an example of the elliptical fitting. We exclude parts of contour lines in the masked region from the fitting. There are \NN clusters that satisfy equation (\ref{eq:cond}) for $R/r_{200}=0.1,0.2,0.3$ and $0.4$. We exclude 7 clusters, Coma, Perseus, A1060, M87, A2256, A3888, and RXJ0658-55 from our analysis because we cannot fit even $R/r_{200} = 0.1 $ due to their redshifts or the virial radius, their position in the field of view or the bullet cluster. The best fit value of $q$ and its error for each cluster are listed in Table \ref{tab:q}. The errors of $q$, $\sigma_q$ is typically a few percent of $q$.
 
\section{Probability Distribution of the Axis Ratio \label{sec:pdf}}

\begin{figure}[!tbh]
  \centerline{\includegraphics[width=85mm]{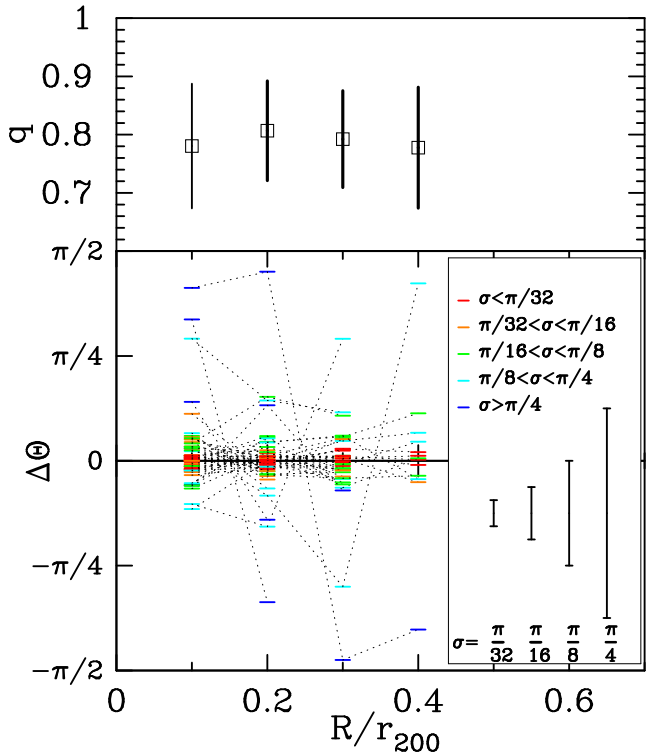}}
  \caption{Radial dependences of the projected axis-ratio and the axis direction. The upper panel shows the average profile of the axis ratio $q$. We average over \NN clusters for $R/r_{200}=0.1,0.2,0.3$ and 0.4, respectively. Black errors indicate the standard deviation for clusters. The Lower panel display the difference of the axis direction, $\Delta \Theta(R_i) \equiv \Theta(R_i)-\overline{\Theta}$. The average axis direction $\overline{\Theta}$ is derived by the average of the unit vector with $\Theta(R_i)$, $\overline{\Theta} = \arccos{[ u_x/(u_x^2 + u_y^2)^{1/2} ]} $, where $u_x = \sum_i \cos{\Theta (R_i)}$ and $u_y = \sum_i \sin{\Theta (R_i)}$. Each dashed line connects $\Delta \Theta$ for different radii of an identical cluster. Each color indicates the standard deviation of $\Theta(R_i)$ for the $j$-$th$ cluster $\sigma_{\Theta,j}$ : red, orange, green, cyan, and blue.   \label{fig:raddep}}
\end{figure}

Figure \ref{fig:raddep} shows the radial dependence of the projected axis ratio and the axis direction. In the upper panel, we plot the average of $q$ for clusters as a function of $R/r_{200}$. Although there is a slight increment of the axis ratio with $z$, the difference between $R/r_{200}=0.1$ and $0.3$ does not significantly change ($\approx 0.04$). The lower panel shows the radial dependence of the difference of $\Theta(R_i)$ around the radial average, $\Delta \Theta(R_i) \equiv \Theta(R_i)-\overline{\Theta}$. The average axis direction $\overline{\Theta}$ is computed from the average of the unit vector with an angle $\Theta(R_i)$, $\overline{\Theta} \equiv \arccos{[ u_x/(u_x^2 + u_y^2)^{1/2} ]} $, where $u_x = \sum_i \cos{\Theta (R_i)}$ and $u_y = \sum_i \sin{\Theta (R_i)}$. For most clusters, the axis directions are aligned within $10$ degree. The points with $\Delta \Theta > \pi/8$ have large errors as displayed by color. These large statistical errors of the axis direction mainly originate from the uncertainty due to their high axis ratio (near spherical). Despite some variations of the axis ratio and the axis direction, we find that the surface brightness contours are well approximated by the ellipses with the same direction and axis ratio for different radii.
 
\begin{figure}[!tbh]
  \centerline{\includegraphics[width=85mm]{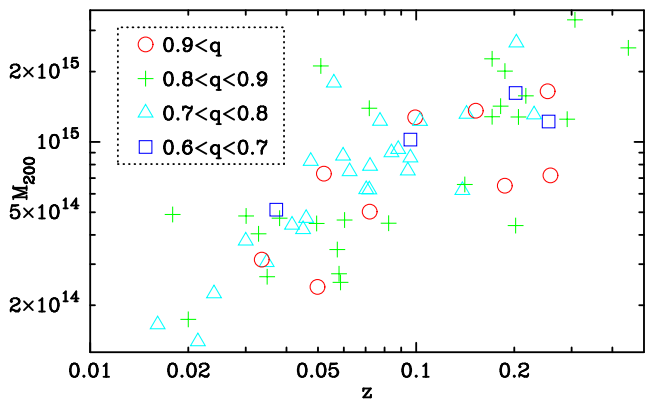}}
  \caption{Scatter plot of clusters with different axis ratio on a $z$-$M_{200}$ plane. Different types of symbols indicate the axis ratio of the cluster: $0.9<q$ (circle), $0.8<q<0.9$ (cross), $0.7<q<0.8$ (triangle), and $0.6<q<0.7$ (square).  \label{fig:zmass}}
\end{figure}

JS02 reported the negative redshift dependence and the (slightly) positive mass dependence on the axis ratio. Figure \ref{fig:zmass} displays the redshift and mass dependence of the axis ratio. Since the present XMM-newton catalogue is heterogeneous, Figure \ref{fig:zmass} does not show any clear dependence of the axis ratio on $z$ or $M_{200}$.

Let us compare the observed axis ratio distribution of X-ray halo with a theoretical expectation. We first construct a simple model of the axis ratio distribution of the X-ray halo based on the triaxial model of the dark matter halo. Our assumptions are as follows, 1)\, The dark matter halo is a homeoid ellipsoid. (has the constant axis ratio and perfect alignment of the axis direction). 2)\, The gas distribution is perfectly governed by the dark matter potential through the hydrostatic equilibrium. In other words, we neglect the self gravity of the gas and regard a cluster as a completely relaxed system. 3)\, The X-ray surface brightness distribution depends on the gas density only, that is, we assume that the cluster is isothermal. Although the actual simulated halo shows a slight radial dependence of the axis ratio as reported by JS02 (see Figure 3 of their paper), we assume the constant axis model. 

JS02 suggested the triaxial density model with three axes $a_1 \ge a_2 \ge a_3$ described as
\begin{eqnarray}
\rho (\tR) = \frac{\rho_0 }{{(\tR/\tR_s)}^\alpha [1+{(\tR/\tR_s)}^{3-\alpha}]}, 
\label{eq:densd}
\end{eqnarray}
where $\alpha=1$ for the cluster size halo (the triaxial NFW profile), $\alpha=1.5$ for the galaxy size halo and $\tR$ is the major axis length: 
\begin{eqnarray}
\tR^2 =  x^2 + y^2/\muB^2 + z^2/\muA^2,
\end{eqnarray}
where $\muA \equiv a_3/a_1$ and $\muB \equiv a_2/a_1$ denote the minor-to-major and medium-to-major axis ratio, respectively. JS02 also provided the fitting formula of the joint probability density function (PDF) of $\muA$ and $\muB$ at $A^{(3)} = 2500$, $p_{\mathrm{DM}} (\muA, \muB; M_{\mathrm{200}}; z) = p_A(\muA; M_{\mathrm{200}}; z) p_B(\muB|\muA)$, where 
\begin{eqnarray}
 p_A(\muA; M_{\mathrm{200}}; z) &=& \frac{1}{\sqrt{2 \pi} \sigma_s} \exp{ \left\{ -\frac{[  \muA  g(M_{\mathrm{200}},z) - 0.54 ]^2}{2 \sigma_s^2} \right\}},  \nonumber \\
 p_B(\muB|\muA) &=& \frac{3}{2(1-\rmin)} \left[ 1 - \left(\frac{2 \muB - 1 -\rmin }{1 - \rmin} \right)^2 \right], \nonumber \\
g(M_{\mathrm{200}},z) &\equiv& \left( \frac{M_{\mathrm{200}}}{M_\ast} \right)^{0.07 \Omega(z)^{0.7}}, 
\label{eq:JSPDF}
\end{eqnarray}
where $M_\ast$ is the characteristic nonlinear mass at $z$ so that the rms top-hat smoothed over density at the scale is 1.68, $\sigma_s=0.113$, and $\rmin = \muA$ for $\muA \ge 0.5$ and $\rmin = 0.5$ for $\muA < 0.5$. The $A_{(3)}$ is $\tR$ with which the ellipsoid having overdensity 2500. Note that $A_(3)=2500$ corresponds to the typical radius $\approx 0.3 \, r_{\mathrm{200}}$. 

Under the hydrostatic equilibrium assumption, the gas isodensity surface is identical to the isopotential surface. The potential of a homeoid at the position $(x,y,z)$ is provided by \citet{1987gady.book.....B},
\begin{eqnarray}
\Phi(x,y,z) &=& - \pi G \left( \frac{a_2 a_3}{a_1} \right) \int_0^\infty \frac{\psi(\infty)-\psi(m)}{\sqrt{(a_1^2 + \tau) (a_2^2 + \tau) (a_3^2 + \tau)}} \, d \tau \nonumber \\
\psi(m) &=& 2 \int_0^m \rho (\tR) \tR d \tR \nonumber \\
m^2 &\equiv&  \frac{x^2}{\tau + 1} + \frac{y^2}{\tau + {\muB}^2} + \frac{z^2}{\tau + {\muA}^2}.
\end{eqnarray}

Assuming the density distribution in equation (\ref{eq:densd}), we obtain 
\begin{eqnarray}
\psi(m) = \frac{2 \tR_s^2 \rho_0}{2 - \alpha} \left( 1 + \frac{\tR_s}{m} \right)^{\alpha-2}.
\end{eqnarray}
We can rewritten the potential as,
\begin{eqnarray}
\Phi(x^{\prime},y^{\prime},z^{\prime}) &=& - 2 \pi G \frac{\tR_s^2 \rho_0}{\alpha - 2} \muA \muB \, \Xi(x^{\prime},y^{\prime},z^{\prime}) \nonumber \\
\Xi(x^{\prime},y^{\prime},z^{\prime}) &\equiv& \int_0^\infty \frac{1 - [1 + (m^{\prime})^{-1}]^{\alpha-2}}{\sqrt{(1 + \tau) (\muB^2 + \tau) (\muA^2 + \tau) }} \, d \tau, \\
\end{eqnarray}
where $x^{\prime},y^{\prime},z^{\prime}$ and $m^{\prime}$ denote $x,y,z$ and $m$ normalized by $\tR_s$, respectively. Although \cite{2003ApJ...585..151L} provided a perturbative approach to solve $\Phi$, we directly derive the axis ratio by numerical integration. The isopotential surface defined by $\Phi(x,y,z)=\mathrm{const}=\Phi(x_\ast,0,0)$ can be  approximated by the triaxial ellipsoid with the minor to major and the medium to major axis ratio, 
\begin{eqnarray}
\muAgas (\muA,\muB;x_\ast^{\prime}) \equiv \frac{z_\ast^{\prime}}{x_\ast^{\prime}} \mbox{, and } \\
\muBgas (\muA,\muB;x_\ast^{\prime}) \equiv \frac{y_\ast^{\prime}}{x_\ast^{\prime}},
\end{eqnarray}
where $y_\ast^{\prime}$ and $z_\ast^{\prime}$ are obtained by solving following relations,
\begin{eqnarray}
\Xi(x_\ast^{\prime},0,0) &=& \Xi(0,0,z_\ast^{\prime}) = \mathrm{const}, \\
\Xi(x_\ast^{\prime},0,0) &=& \Xi(0,y_\ast^{\prime},0) = \mathrm{const}.
\end{eqnarray}
Thus, we regard that the gas density distribution as an ellipsoid with constant axis ratios $\muAgas$ and $\muBgas$. 

One uncertainty of this model is the choice of $x_\ast^{\prime}$. Because the fitting formula given in equation (\ref{eq:JSPDF}) is based on $\tR \approx 0.3 \, r_{200}$, it is natural to adopt $x_\ast^{\prime} = 0.3 \, r_{200}/\tR_s = 0.3 \, c$, where $c$ is the concentration parameter. For simplicity, we assume the typical value $c=3$ found in simulations (JS02). Because most contour lines we investigate are below $0.3 \, r_{200}$, we also consider $x_\ast^{\prime}=0.3 \,,0.6$ which approximately corresponds to the contours $i=1,2$ in addition to $0.9$ ($i=3$). 

\begin{figure}[tbh]
  \centerline{\includegraphics[width=85mm]{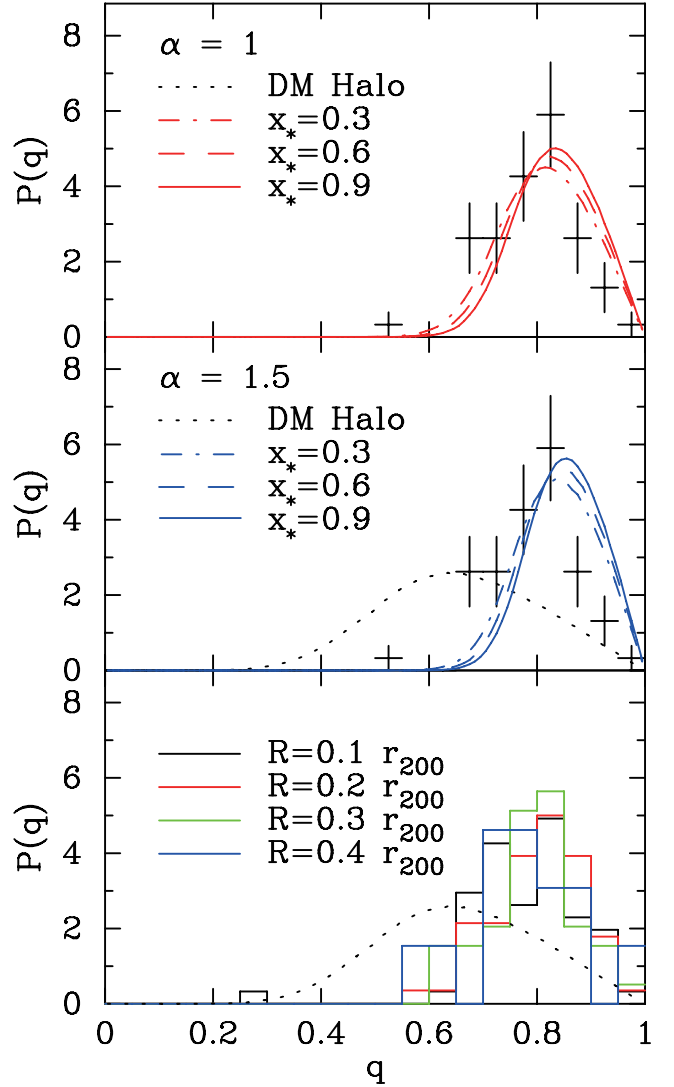}}
  \caption{Comparison the theoretical prediction of the axis ratio distribution (lines) with observation (error bars). We use {\it the radial averaged} axis ratio of each cluster. Error bars are Poissonian. Lines in the top and middle panels are theoretical curves with the density profile (Eq.[\ref{eq:densd}]) with $\alpha=1$ and $\alpha=1.5$. Solid, dashed, and dot-dashed lines indicate $x_\ast^\prime=0.9, 0.6$ and $0.3$, respectively. For reference, we plot the axis ratio distribution of the underlying dark matter halo, which is expected from the lensing observation for instance. The bottom panel displays the PDF for different radii. Black, red, green, and blue lines indicate $R=0.1, 0.2, 0.3$ and $0.4 r_\mathrm{200}$, respectively. \label{fig:sch}}
\end{figure}

The axis ratio of the ellipse made by the projection of the ellipsoid is obtained by \citep{1985MNRAS.212..767B}  
\begin{eqnarray}
&\,& \hat{q} (\theta,\phi; \muAgas, \muBgas) =\sqrt{\frac{A+C-\sqrt{(A-C)^2+B^2}}{A+C+\sqrt{(A-C)^2+B^2}}} \nonumber \\
A &\equiv& \muAgas^{-2} \cos^2{\theta} (\sin^2{\phi} + \muBgas^{-2} \cos^2{\phi}) + \muBgas^{-2} \sin^2{\theta} \nonumber \\
B &\equiv& \muAgas^{-2} ( 1 - \muBgas^{-2} )  \cos{\theta} \sin{2 \phi} \nonumber \\
C &\equiv& \muAgas^{-2} (\muBgas^{-2} \sin^2{\phi} + \cos^2{\phi}),
\end{eqnarray}
where $\theta$ and $\phi$ are the polar coordinates of the line of sight. The PDF of the projected axis ratio is given by
\begin{eqnarray}
&\,& p(q; x_\ast^{\prime}; M_{\mathrm{200}}; z) = \frac{1}{4 \pi} \int_0^{2 \pi} d \phi \int_0^{\pi} \sin{\theta} d \theta \int^1_{\muB} d \muA \int^1_0 d \muB   \nonumber \\
&\times& \delta_{D}(q - \hat{q} (\theta,\phi; \muAgas (\muA,\muB;x_\ast^{\prime}) , \muAgas (\muA,\muB;x_\ast^{\prime}) ) ) \nonumber \\ 
&\times& p_{\mathrm{DM}} (\muA, \muB; M_{\mathrm{200}}; z),
\end{eqnarray}
where $\delta_{D}(x)$ is the delta function. As shown in Figure \ref{fig:zmass}, the cluster sample is not homogeneous for mass and redshift. We do not know the mass or redshift selection function of the catalogue. Therefore we use the ensemble averaged PDF, not using the fixed mass and redshift,
\begin{eqnarray}
P (q; x_\ast^{\prime}) \equiv \frac{1}{N} \sum_{i=1}^{N} p(q; x_\ast^{\prime}; M_{\mathrm{200},i}; z_i),
\label{eq:theoex}
\end{eqnarray}
 for comparison with observation. We adopt the virial mass and redshift of each cluster for $M_{\mathrm{200},i}$ and $z_i$.

Figure \ref{fig:sch} displays the observed axis ratio distribution and the theoretical prediction described in equation (\ref{eq:theoex}). Because the radial dependence of the ellipse is relatively small, we adopt the radial average for each cluster as the observed axis ratio. We plot the theoretical prediction described in equation (\ref{eq:theoex}) by performing Monte Carlo methods. The number of realization for each cluster is $128^2$. The top and bottom panels show the results for $\alpha=1.0$ and $1.5$, respectively. We find that both models significantly agree with observational result although the NFW triaxial model is slightly better than that of $\alpha=1.5$. We also confirm that the model uncertainty due to the selection of $x_\ast^\prime$ is relatively small. The bottom panel displays the PDFs for the different radii $R=0.1,0.2,0.3,$ and $0.4 r_\mathrm{200}$. There is no strong dependence of the PDF on the radius between $0.1 - 0.4 r_\mathrm{200}$

Because our model is quite idealistic, many sources of the systematic errors are possible. For instance, the deviation from the isothermality affects the relation between the gravitational potential and X-ray surface brightness. Under the hydrostatic equilibrium assumption, the contour of the gravitational potential is identical to $n T$ and $n ^2 \Lambda(T)$. Let us consider two points A and B on the isopotential surface. The relation of the local X-ray emissivity at A ($\epsilon_A$) and B ($\epsilon_B$) is $\epsilon_B = [T_A^2 \Lambda (T_B) / (T_B^2 \Lambda (T_A))] \epsilon_A \approx (T_A/T_B)^{3/2} \epsilon_A$, where $T_A$ and $T_B$ are temperature, $n_A$ and $n_B$ are gas density at A and B, and $\Lambda (T)$ is the X-ray emissivity. If assuming the $\beta$ model, the surface brightness profile follows $\propto r^{-3 \beta}$ at out of the core. Then we approximately estimate the change of gas emissivity contour from as $\Delta \sim (T_B/T_A)^{3/4}$. Therefore, 10 \% global deviation of temperature from isopotencial surface corresponds to $\lesssim 10 $ \% change of the iso-emissivity surface or the axis ratio of X-ray haloes. 
 We also note that our model use the shape of the potential at a given radius and ignore the projection effect of the gas density. This assumption is also a possible source of systematics.  Moreover, the hydrostatic equilibrium assumption itself may not valid for merger clusters. A rich variety of structural complexity seen by X-ray observations like shock fronts \citep[e.g.,][]{2002ApJ...567L.115J}, cold fronts \citep[e.g.,][]{2000ApJ...541..542M}, X-ray holes \citep[e.g.,][]{2002MNRAS.331..369F}, gas fluctuations \citep[e.g.,][]{2008ApJ...687..936K}, and AGN feedback are also possible sources of the systematic error of the theoretical prediction. Considering the simplicity of our current model, however, the agreement is remarkable.

\subsection{Comparison with Previous Studies}

In this section, we compare our results with other studies by Einstein, ROSAT, and Chandra. \cite{1995ApJ...447....8M} provided the distribution of the axis ratio for 51 clusters observed by Einstein. The mean axis ratio of their data  ($\langle q \rangle = 0.80$) is similar to our results ($\langle q \rangle \equiv \sum_{j=1}^{\Nsample} (\sum_{i=1}^{i_{\mathrm{max}}} q_j(c_i)/i_{\mathrm{max}})/\Nsample = 0.791 \pm 0.004$), where $\Nsample=62$ is the number of the clusters and $i_{\mathrm{max}}$ is the number of contour lines for each cluster. While the dispersion of their data ($s_q = 0.12$) is slightly larger than ours  ($s_q = 0.095 \pm 0.007$), it is likely that the spatial resolution and sensitivity between Einstein and XMM or the different fitting methods make a difference of the dispersion. 

\cite{2007MNRAS.377..883F} discussed an average ellipticity ($\epsilon \equiv 1 - q^2$) and its dispersion by the flux-limited samples of ROSAT clusters. They provided the mean ellipticity $\langle \epsilon \rangle = 0.376 \pm 0.019$ and its dispersion $s_\epsilon = 0.122 \pm 0.014$.  Although the mean ellipticity of our analysis ($\langle \epsilon \rangle \equiv 1 - \langle q^2 \rangle = 0.365 \pm 0.007$) is slightly smaller than their result, the dispersion ($s_\epsilon = 0.144 \pm 0.005$) is good agreement.   \cite{2007A&A...467..485H} also derived the mean ellipticity $\epsilon^{\prime} = 0.22$ with 101 clusters observed by Chandra (their ellipticity is defined by $\epsilon^{\prime} \equiv 1 - q$) . Converting to the mean axis ratio, we obtain $\langle q \rangle = 0.78$. \cite{2005ApJ...625..108D} also provided the mean axis ratio $\langle q \rangle = 0.81$ for 25 clusters by Chandra and XMM-Newton. These two results are consistent with ours ($\langle q \rangle = 0.791 \pm 0.004$).

In this paper, we compute the theoretical prediction of the PDF based on the phenomenological model of JS02 under the hydrostatic equilibrium and isothermal assumptions, while, \cite{2007MNRAS.377..883F} derived the mean ellipticity and dispersion based on 46 haloes in the N-body simulation. Their prediction of the mean ellipticity is $\langle \epsilon \rangle = 0.323 \pm 0.013$ and $s_\epsilon = 0.138 \pm 0.008$. In order to compare them with our theoretical prediction, we compute the theoretical value for the same cosmological parameter of their N-body simulation ($\Omega_m=0.3, \lambda=0.7, \sigma_8=0.9, h=0.7$). because they used 46 haloes with $M=(1-4) \times 10^{14} h^{-1}  M_{\odot}$, we choose two cases of halo mass at $z=0$, $M_1= 10^{14} h^{-1} M_\odot$ and $M_4= 4 \times 10^{14} h^{-1} M_\odot$. Performing the Monte Carlo simulation ($n=10^5$) with $\alpha=1$ and $x_\ast^\prime=0.3$, we obtain $\langle \epsilon \rangle = 0.307$ and $s_\epsilon = 0.128$ for $M=M_1$ or $\langle \epsilon \rangle = 0.320$ and $s_\epsilon = 0.130$ for $M=M_4$. Therefore our model is consistent with their model.

In short summary, our results of the mean axis ratio and its dispersion for the observation and the theoretical prediction are basically consistent with previous studies except for the double peak feature seen in the result by Einstein. More detailed comparison needs to reduce uncertainty due to the fitting methods or the selection effects.

\begin{table*}[!tbh]
\begin{center}
\caption{Axis ratio $q(R_i)$ and its standard deviation $\sigma_q(R_i)$ for the threshold of the contours $R_i$. \label{tab:q}}
  \begin{tabular}{lcccccccc}
   \hline\hline
 Cluster  & $q(R_1)$ & $\sigma_{q}(R_1)$ &  $q(R_2)$ & $\sigma_{q}(R_2)$ &  $q(R_3)$ & $\sigma_{q}(R_3)$ &  $q(R_4)$ & $\sigma_{q}(R_4)$ \\
   \hline
2A0335+096 & 0.84 & 0.004 & 0.84 & 0.01 & 0.79 & 0.02 & - & - \\
A13 & 0.7 & 0.06 & 0.76 & 0.05 & 0.75 & 0.07 & - & - \\
A68 & 0.69 & 0.1 & 0.67 & 0.04 & 0.73 & 0.05 & - & - \\
A85 & 0.94 & 0.02 & 0.83 & 0.02 & - & - & - & - \\
A133 & 0.86 & 0.03 & 0.84 & 0.02 & 0.8 & 0.05 & - & - \\
A209 & 0.65 & 0.06 & 0.8 & 0.05 & 0.79 & 0.05 & 0.79 & 0.09 \\
A262 & 0.8 & 0.02 & - & - & - & - & - & - \\
A383 & 0.91 & 0.03 & 0.93 & 0.03 & - & - & - & - \\
A399 & 0.89 & 0.04 & 0.91 & 0.04 & 0.69 & 0.03 & - & - \\
A400 & 0.67 & 0.05 & - & - & - & - & - & - \\
A478 & 0.74 & 0.005 & 0.77 & 0.01 & 0.84 & 0.01 & 0.88 & 0.03 \\
A496 & 0.82 & 0.01 & 0.77 & 0.02 & - & - & - & - \\
A576 & 0.81 & 0.04 & 0.89 & 0.03 & - & - & - & - \\
A665 & 0.73 & 0.04 & 0.82 & 0.04 & 0.94 & 0.03 & 0.96 & 0.04 \\
A773 & 0.72 & 0.04 & 0.88 & 0.05 & 0.77 & 0.05 & - & - \\
A1068 & 0.7 & 0.03 & 0.72 & 0.03 & 0.75 & 0.06 & - & - \\
A1413 & 0.74 & 0.03 & 0.68 & 0.02 & 0.74 & 0.03 & 0.78 & 0.04 \\
A1589 & 0.66 & 0.11 & 0.67 & 0.06 & 0.62 & 0.05 & - & - \\
A1650 & 0.73 & 0.01 & 0.78 & 0.02 & 0.82 & 0.03 & 0.73 & 0.05 \\
A1689 & 0.9 & 0.03 & 0.87 & 0.02 & 0.84 & 0.04 & 0.82 & 0.04 \\
A1775 & 0.81 & 0.02 & 0.94 & 0.02 & 0.87 & 0.07 & - & - \\
A1795 & 0.76 & 0.004 & 0.79 & 0.01 & 0.76 & 0.01 & - & - \\
A1835 & 0.94 & 0.02 & 0.9 & 0.02 & 0.91 & 0.03 & - & - \\
A1837 & 0.7 & 0.01 & 0.64 & 0.04 & - & - & - & - \\
A1914 & 0.75 & 0.02 & 0.85 & 0.03 & 0.88 & 0.04 & - & - \\
A1991 & 0.8 & 0.03 & 0.8 & 0.03 & - & - & - & - \\
A2029 & 0.76 & 0.01 & 0.83 & 0.02 & 0.84 & 0.03 & - & - \\
A2052 & 0.79 & 0.01 & 0.83 & 0.01 & - & - & - & - \\
A2065 & 0.8 & 0.02 & 0.71 & 0.02 & 0.64 & 0.02 & 0.63 & 0.04 \\
A2163 & 0.65 & 0.06 & 0.88 & 0.04 & 0.74 & 0.03 & 0.72 & 0.03 \\
A2199 & 0.81 & 0.01 & - & - & - & - & - & - \\
A2204 & 0.9 & 0.01 & 0.87 & 0.02 & 0.83 & 0.03 & - & - \\
A2218 & 0.8 & 0.05 & 0.85 & 0.03 & 0.81 & 0.04 & - & - \\
A2319 & 0.7 & 0.01 & 0.77 & 0.02 & - & - & - & - \\
A2589 & 0.73 & 0.01 & 0.69 & 0.03 & - & - & - & - \\
\hline
\end{tabular}
\end{center}
\end{table*}

\begin{table*}[!tbh]
\begin{center}
\caption{continued.}
  \begin{tabular}{lcccccccc}
   \hline\hline
   & $q(R_1)$ & $\sigma_{q}(R_1)$ &  $q(R_2)$ & $\sigma_{q}(R_2)$ &  $q(R_3)$ & $\sigma_{q}(R_3)$ &  $q(R_4)$ & $\sigma_{q}(R_4)$ \\
   \hline
A2597 & 0.82 & 0.01 & 0.79 & 0.01 & 0.85 & 0.02 & 0.7 & 0.04 \\
A2626 & 0.83 & 0.03 & 0.88 & 0.02 & 0.77 & 0.04 & - & - \\
A2667 & 0.73 & 0.03 & 0.83 & 0.03 & 0.81 & 0.05 & - & - \\
A2717 & 0.98 & 0.02 & 1.0 & 0.02 & - & - & - & - \\
A3112 & 0.76 & 0.01 & 0.79 & 0.02 & 0.76 & 0.03 & - & - \\
A3158 & 0.72 & 0.03 & 0.79 & 0.02 & 0.67 & 0.02 & - & - \\
A3558 & 0.67 & 0.01 & 0.77 & 0.01 & - & - & - & - \\
A3560 & 0.91 & 0.05 & 0.92 & 0.05 & - & - & - & - \\
A3581 & 0.81 & 0.01 & 0.79 & 0.07 & - & - & - & - \\
A3827 & 0.9 & 0.02 & 0.93 & 0.03 & 0.91 & 0.03 & - & - \\
A3911 & 0.27 & 0.12 & 0.59 & 0.02 & 0.64 & 0.02 & - & - \\
A3921 & 0.73 & 0.02 & 0.73 & 0.02 & 0.67 & 0.02 & 0.59 & 0.02 \\
A4059 & 0.76 & 0.01 & 0.74 & 0.02 & - & - & - & - \\
AWM7 & 0.69 & 0.01 & - & - & - & - & - & - \\
E1455+2232 & 0.9 & 0.02 & 0.9 & 0.04 & 0.88 & 0.05 & - & - \\
EXO0422 & 0.9 & 0.01 & 0.88 & 0.03 & - & - & - & - \\
Hydra & 0.94 & 0.03 & 0.89 & 0.03 & 0.78 & 0.03 & 0.78 & 0.04 \\
Klemola44 & 0.71 & 0.01 & 0.66 & 0.01 & - & - & - & - \\
MKW3S & 0.7 & 0.01 & 0.74 & 0.02 & 0.73 & 0.02 & - & - \\
MKW4 & 0.83 & 0.05 & - & - & - & - & - & - \\
PKS0745-19 & 0.76 & 0.01 & 0.83 & 0.02 & 0.77 & 0.03 & - & - \\
RXCJ0605.8-3518 & 0.88 & 0.03 & 0.72 & 0.03 & 0.86 & 0.05 & - & - \\
RXJ1347-1145 & 0.73 & 0.02 & 0.84 & 0.03 & 0.84 & 0.04 & 0.85 & 0.05 \\
Sersic159-3 & 0.82 & 0.01 & 0.83 & 0.01 & 0.83 & 0.03 & - & - \\
Triangulum & 0.83 & 0.01 & 0.7 & 0.02 & - & - & - & - \\
ZW3146 & 0.81 & 0.01 & 0.85 & 0.01 & 0.98 & 0.03 & 0.89 & 0.04 \\
\hline
\hline
   & $\overline{q}(R_1)$ & $\sigma_{\overline{q}}(R_1)$ &  $\overline{q}(R_2)$ & $\sigma_{\overline{q}}(R_2)$ &  $\overline{q}(R_3)$ & $\sigma_{\overline{q}}(R_3)$ &  $\overline{q}(R_4)$ & $\sigma_{\overline{q}}(R_4)$ \\
\hline
 & 0.78 & 0.005 & 0.81 & 0.004 & 0.79 & 0.006 & 0.78 & 0.013\\
\hline
\end{tabular}
\end{center}
\end{table*}

\section{Conclusion and Summary}
In this paper, we have examined the axis ratio distribution of X-ray halos using the XMM-Newton cluster catalogue. By fitting the surface brightness contours by ellipses, we confirmed that the radial dependence of the axis ratio and the axis direction is relatively small, that is, the X-ray halo is well approximated by the triaxial ellipsoid. Constructing the simple model based on the hydrostatic equilibrium of the underlying triaxial dark matter halo proposed by  \citet{2002ApJ...574..538J}, we found that the observed PDF of the axis ratio of X-ray halo agrees well with the theoretical prediction.  We have shown that the axis ratio distribution of a sample of X-ray clusters predicted by the $\Lambda$ CDM model agrees with that measured by XMM-Newton, consistent with the findings of previous studies using different data sets, which is encouraging given the simplifying assumptions adopted in all such model comparisons, including our own.

\acknowledgments 
We are deeply grateful to Akio Hoshino for his instruction of the X-ray analysis. 
We thank Takahiro Nishimichi, Thierry Sousbie, and Yasushi Suto for 
useful discussions. We also thank an anonymous referee for his constructive comments. 
HK is supported by a JSPS (Japan Society for Promotion of Science) Grant-in-Aid for science fellows. This work is also supported by Grant-in-Aid for Scientific research from JSPS and from the Japanese Ministry of Education,
Culture, Sports, Science and Technology (Nos. 20$\cdot$10466 and 22$\cdot$5467), and by the
JSPS Core-to-Core Program ``International Research Network for Dark
Energy''.


\newpage


\end{document}